\documentclass{PoS}

\newcommand{\bk}{{\bf{k}}}
\newcommand{\bl}{{\bf{l}}}

\newcommand{\bq}{{\bf{q}}}

\newcommand{\bkappa}{{\bf{\kappa}}}

\newcommand{\aem}{\mbox{$\alpha_{\rm{em}}$}}

\title{Forward production of Drell-Yan dileptons at high energies and low
dilepton invariant masses in a $k_t$-factorization approach: Do we see onset
of saturation?}

\ShortTitle{Forward production of Drell-Yan dileptons at high energies in a $k_t$-factorization approach}

\author{\speaker{Antoni Szczurek}%
        \thanks{
   supported by the Polish National Science 
       Centre grant DEC-2014/15/B/ST2/02528.     
                }\\
       Institute of Nuclear Physics\\
       E-mail: \email{antoni.szczurek@ifj.edu.pl}}
      
\author{Wolfgang Sch\"afer\\
       Institute of Nuclear Physics, Krakow\\
       E-mail: \email{wolfgang.schafer@ifj.edu.pl}}

\abstract{
We discuss Drell-Yan production of dileptons at high energies 
in forward rapidity region in a hybrid high-energy approach.
This approach uses unintegrated gluon distributions in one proton 
and collinear quark/antiquark distributions in the second proton.
Corresponding momentum-space formula for the differential cross sections 
in high-energy approximation has been derived.
The relation to the commonly used dipole approach is discussed.
We conclude and illustrate that some results of the dipole approaches 
are too approximate, as far as kinemtics is considered, and in fact 
cannot be used for real experiments.
We find that the dipole formula is valid only in very forward/backward
rapidity regions ($|y| >$ 5).
Some differential cross sections for
low-mass dilepton production are shown and compared
to the LHCb and ATLAS experimental data.
In distinction to dipole approaches, we include four
Drell-Yan structure functions (the impact of interference
structure functions is rather small for typical experimental
cuts).
We find that both side contributions ($g q/\bar q$ and $q/\bar q g$) 
have to be included even for the LHCb rapidity coverage which
is in contradiction with what is usually done in the dipole approach.
We present results for different unintegrated gluon distributions from
the literature (some of them include saturation effects).
We see no clear hints of saturation even at small $M_{ll}$.}

\FullConference{XXIV International Workshop on Deep-Inelastic Scattering and Related Subjects\\
		11-15 April, 2016\\
		DESY Hamburg, Germany}

\begin{document}

\section{Introduction}

It was proposed some time ago that the Drell-Yan production of low
invariant masses of dileptons in forward directions is 
a good place in searching for the onset of (gluon) saturation 
\cite{Gelis:2002fw}.
Recently a numbers of papers addressing the Drell-Yan process at the LHC appeared,
which use the dipole approach (see e.g. \cite{Ducati:2013cga,GolecBiernat:2010de,Basso:2015pba,Motyka:2014lya}) 
in which the main ingredient is 
the dipole-nucleon cross section parametrized as a function 
of dipole size ($\rho$) and collision energy or a similar equivalent kinematical variable.
 
In the dipole picture only global kinematic variables of the lepton
pair (invariant mass, rapidity and transverse momentum of 
the pair) are used. In contrast, in real experiments one imposes
cuts on pseudorapidities and transverse momenta of leptons.
A real comparison of the theoretical predictions and experimental data
is therefore not possible.

Here we present an alternative formulation in the momentum space 
proposed recently \cite{Schafer:2016qmk}.
This approach allows for explicit treatment of momenta of 
individual leptons ($e^+ e^-$ or $\mu^+ \mu^-$) and 
therefore a comparison to existing experimental data.

\begin{figure}[!ht]
\includegraphics[width=.22\textwidth]{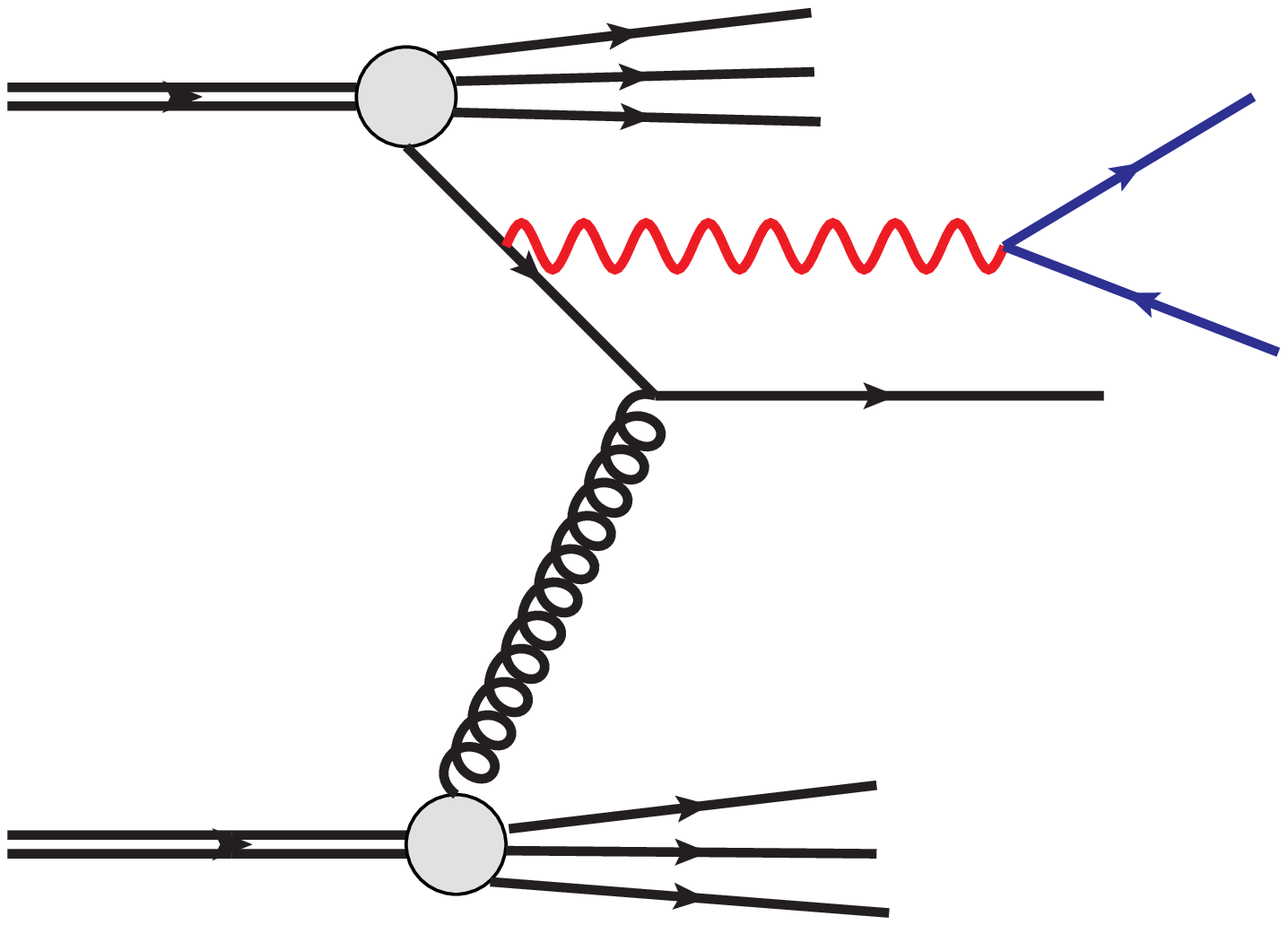}
\includegraphics[width=.22\textwidth]{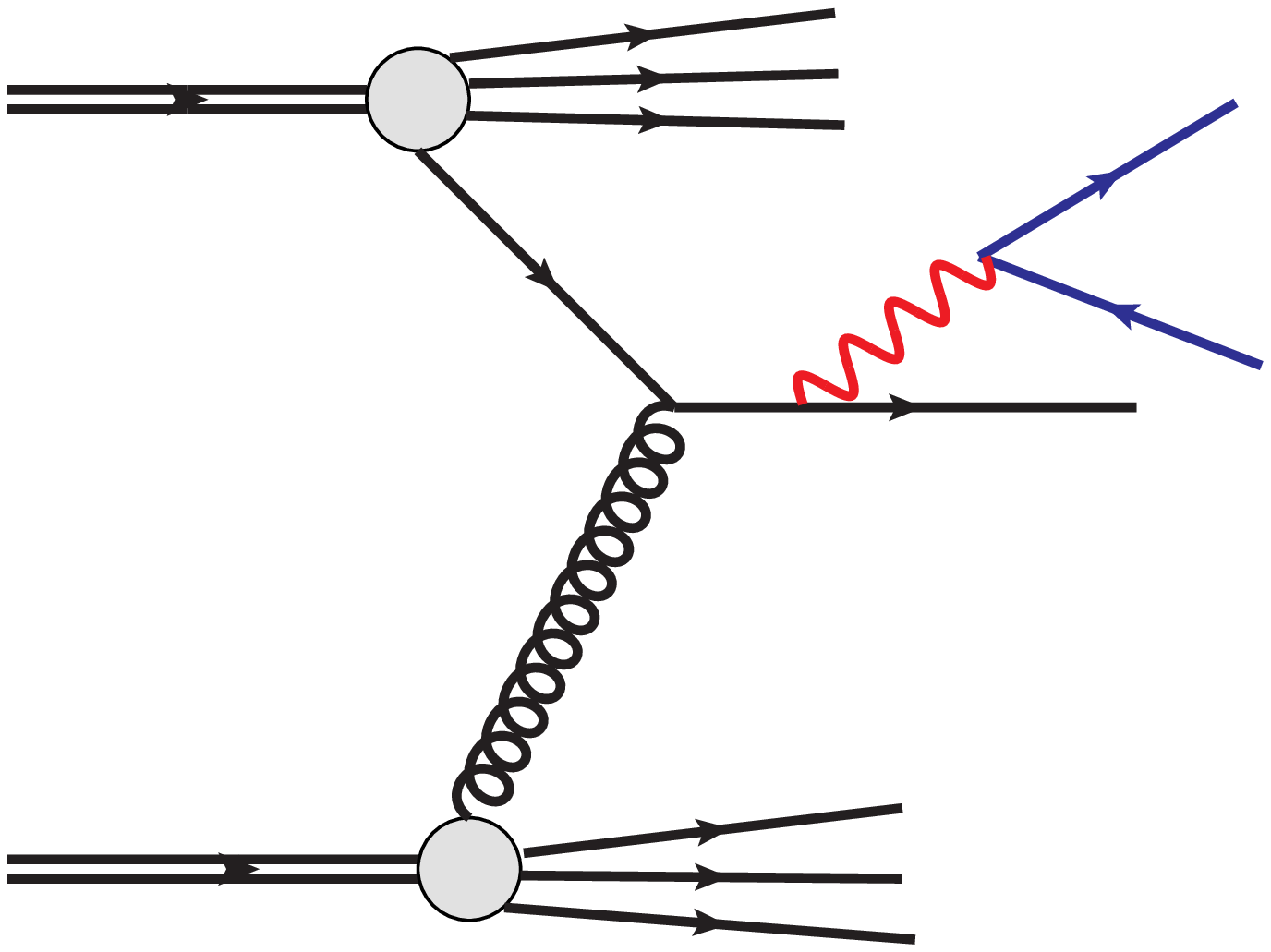}
\includegraphics[width=.22\textwidth]{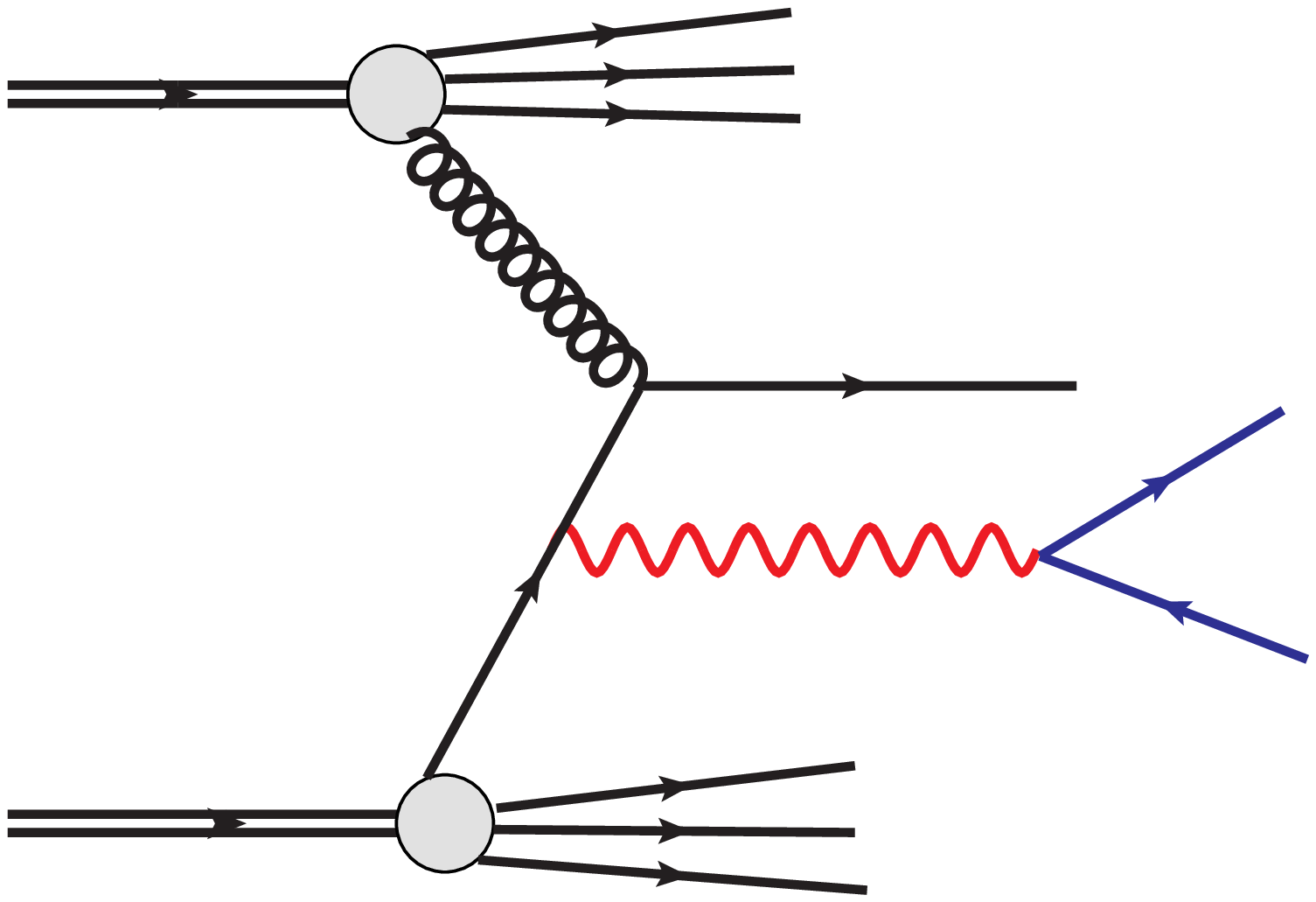}
\includegraphics[width=.22\textwidth]{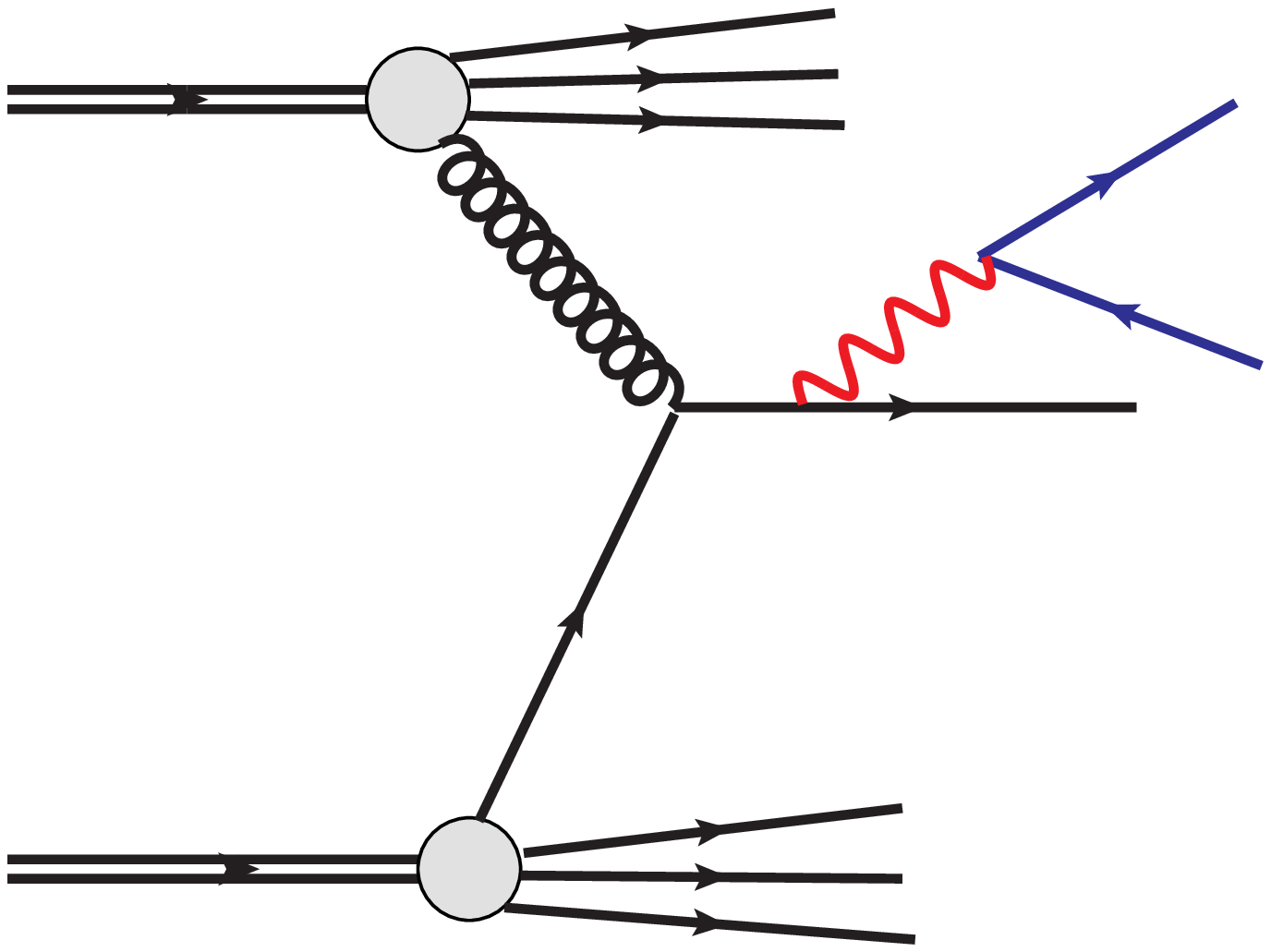}
\caption{\label{fig:diagrams}
The diagrams relevant for forward and backward production of dilepton pairs.
}
\end{figure}

The mechansims considered are shown in Fig.\ref{fig:diagrams}.
%%%%%%%%%%
%%%%%%%%%%%
\section{Formalism and Results}

The inclusive cross section for lepton pair production can be written in the form:
\begin{eqnarray}
{d\sigma(pp \to l^+ l^- X) \over dx_+ dx_- d^2\bk_+ d^2\bk_-} = &&{ \aem \over (2 \pi)^2 M^2} {x_F \over x_+ x_-} \Big\{ \Sigma_T(x_F,\bq,M^2)  D_T\Big({x_+ \over x_F} \Big)  +
\Sigma_L(x_F,\bq,M^2)  D_L\Big({x_+ \over x_F}\Big) \nonumber \\
&& + \Sigma_\Delta(x_F,\bq,M^2)  D_\Delta\Big({x_+ \over x_F}\Big)  \Big({\bl \over |\bl|} \cdot {\bq \over |\bq|} \Big) \nonumber \\
&& +  \Sigma_{\Delta \Delta}(x_F,\bq,M^2) D_{\Delta \Delta}\Big({x_+
  \over x_F}\Big) \Big(2  \Big({\bl \over |\bl|} \cdot {\bq \over |\bq|}
\Big)^2 - 1 \Big) \Big\} \; .
\end{eqnarray}
Here $x_\pm$ are longitudinal (lightcone-) momentum fractions of leptons, $\bk_\pm$ are their transverse momenta.
The heavy virtual photon carries the longitudinal momentum fraction $x_F = x_+ + x_-$ and transverse momentum $\bq = \bk_+ + \bk_-$.
There also appears the light-front relative transverse momentum of $l^+$ and $l^-$:
%%%
\begin{eqnarray}
 \bl = {x_+ \over x_F} \bk_- - {x_- \over x_F} \bk_+ \, .
\end{eqnarray}
%%%
The functions $\Sigma_i(x_F,\bq,M^2),\, i = T,L,\Delta,\Delta\Delta$ correspond to the four helicity structure functions \cite{Oakes:1966}
of inclusive lepton pair production. They contain all information of strong dynamics in the production
of the virtual photon. The functions $D_i$ and the momentum structures in brackets represent the density matrix of decay
of the massive photon into $l^+ l^-$. See \cite{Schafer:2016qmk} for explicit expressions.
Let us concentrate on, say the first two diagrams of Fig. \ref{fig:diagrams}. Then we are dealing with a process where
a fast quark from one proton radiates a virtual photon interacting with a small-$x$ gluon of the other proton.
It is therefore natural to adopt a factorization which involves the collinear quark distribution from one side
and the $k_T$-dependent unintegrated gluon distribution from the other side.
We can write for the functions $\Sigma_i$:
\begin{eqnarray}
\Sigma_i (x_F,\bq,M) &=& \sum_f \int dx_1 dz \, \delta(x_F - z x_1) \,  
\Big[q_f(x_1,\mu^2) + \bar q_f(x_1,\mu^2 )\Big] 
\hat \Sigma_i(z,\bq,M^2) \, .   \nonumber \\
&=& \sum_f  {e_f^2 \alpha_{\rm em} \over 2 N_c} \int_{x_F}^1  dx_1 \,
\Big[q_f(x_1,\mu^2) + \bar q_f(x_1,\mu^2 )\Big]  
%\nonumber \\
%&\times&
\int {d^2 \bkappa \over \pi \bkappa^4} 
 {\cal{F}}(x_2,\bkappa^2) 
\alpha_S(\bar q^2) I_i \Big( {x_F \over x_1} ,\bq,\bkappa \Big) \, . 
\nonumber\\
\end{eqnarray}
Here we have introduced the parton-level functions $\hat \Sigma_i$ which correspond to the process 
$q p \to q \gamma^* p$. In the last line we have given their impact-factor representation 
in terms of the unintegrated gluon distribution ${\cal{F}}(x_2,\bkappa^2)$ and impact factors $I_i$
which can be found in \cite{Schafer:2016qmk}.

An important comment on the longitudinal momentum fractions $x_1, x_2$ is in order.
They must be obtained from the full $l^+ l^- q$ final state:
{\small
\begin{eqnarray}
x_1 &=& \sqrt{ \bk_+^2  \over S} e^{y_+} + \sqrt{ \bk_-^2 \over S} e^{y_-} + 
{\sqrt{ \bk_q^2 \over S} e^{y_q}} \; , \nonumber \\
x_2 &=& \sqrt{ \bk_+^2 \over S} e^{-y_+} + \sqrt{ \bk_-^2  \over S} e^{-y_-} + 
{\sqrt{ \bk_q^2  \over S} e^{-y_q}} \; .
\label{x1_x2}
\end{eqnarray}
}
Here the contribution from the final state (anti-)quark must not be neglected!

%------------------------------------------------------------------------
\begin{figure}[!ht]
\includegraphics[width=7cm]{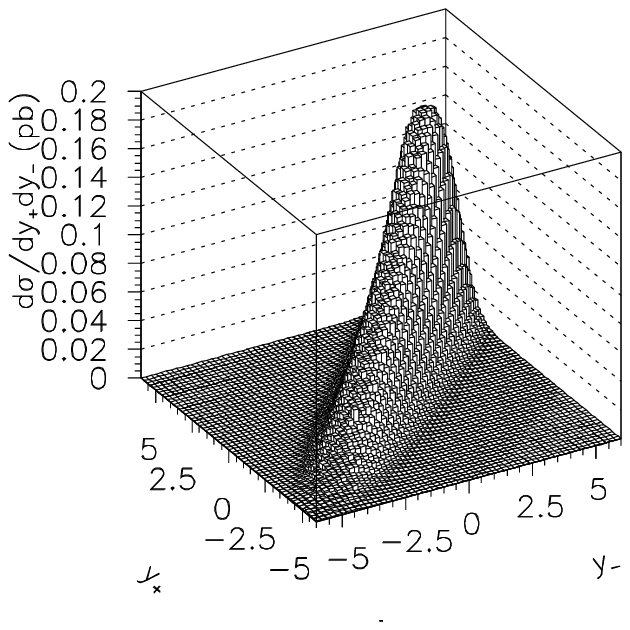}
\includegraphics[width=7cm]{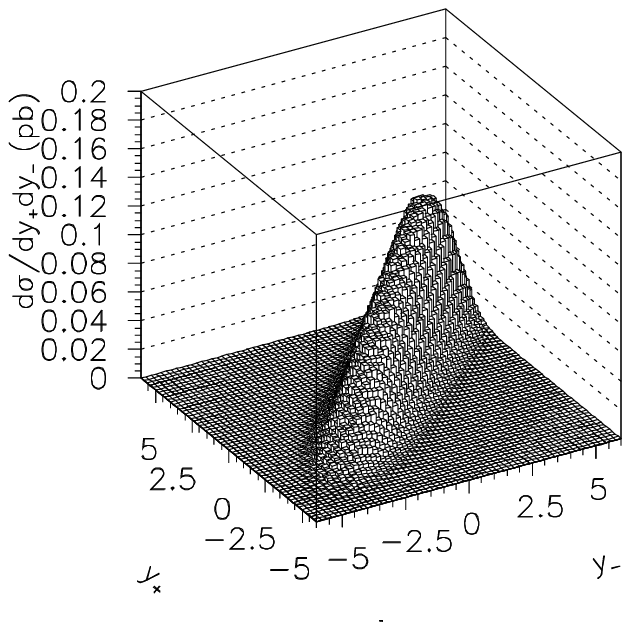}
\caption{\label{fig:full_map_ypym}
Two-dimensional $(y_{+},y_{-})$
distribution for $\sqrt{s}$ = 7 TeV and $k_{T+},k_{T-} >$ 3 GeV
for MSTW08 PDF and KMR (left) and KS (right) UGDFs.
}
\end{figure}
%%%

%------------------------------------------------------------------------
\begin{figure}[!ht]
\includegraphics[width=.5\textwidth]{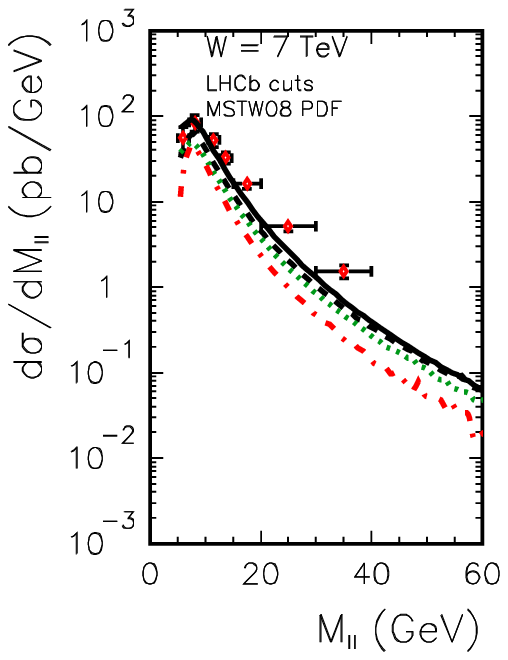}
\includegraphics[width=.5\textwidth]{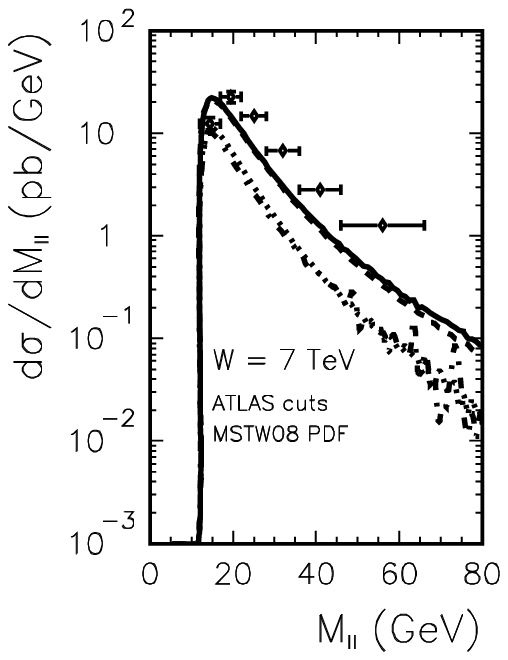}
  \caption{\label{fig:lhcb_dsig_dMll_ugdfs}
Left panel: Invariant mass distribution (only the dominant component)
for the LHCb cuts: 2 $< y_{+},y_{-} <$ 4.5, $k_{T+},k_{T-} >$ 3 GeV
for different UGDFs: KMR (solid), Kutak-Stasto (dashed), AAMS (dotted)
and GBW (dash-dotted). Right panel:  the same for the ATLAS kinematics:
-2.4 $< y_{+},y_{-} <$ 2.4, $k_{T+}, k_{T-} >$ 6 GeV.
Here both $g q/\bar q$ and $q/\bar q g$ contributions have been included.
}
\end{figure}
%------------------------------------------------------------------------

We come to a selection of results presented in \cite{Schafer:2016qmk}.
In Fig. \ref{fig:full_map_ypym} we show a map of the $(y_+,y_-)$ rapidity
plane of leptons for two different unintegratde gluon distributions. 
We see a strong correlation along the diagonal, which is 
a property of the $\gamma^* \to l^+ l^-$ ``decay''. Notice the peak in the 
forward rapidity region -- still there is a nonlegligible amount of radiation into
negative rapidities. The presence of such radiation in the ``backward'' region 
makes it necessary to include all of the diagrams of Fig. \ref{fig:diagrams}.

In Fig. \ref{fig:lhcb_dsig_dMll_ugdfs} we compare our results to experimental data
on so-called ``low-mass'' Drell-Yan.
In the left panel we compare our results to the date from the LHCb collaboration \cite{LHCb-CONF_2012},
which cover the forward rapidity region. We see that a reasonable description of
data can be obtained by an unintegrated gluon distribution constructed by the KMR prescription.
Gluon distributions which include gluon saturation effects do not lead to a good
agreement, certainly there is no trace that gluon saturation would be required 
by the data.
In the right panel, we compare our results to the ATLAS data \cite{Aad:2014qja}.
These are obtained in the central rapidity region, and we see that we are underpredicting the
cross section at larger masses. This is a sign that the aprroach used here is in fact not adequate
for the central rapidity region. For example, here one should also include transverse momenta
of quarks in a consistent manner.

%------------------------------------------------------------------------
\begin{figure}[!ht]
\includegraphics[width=8cm]{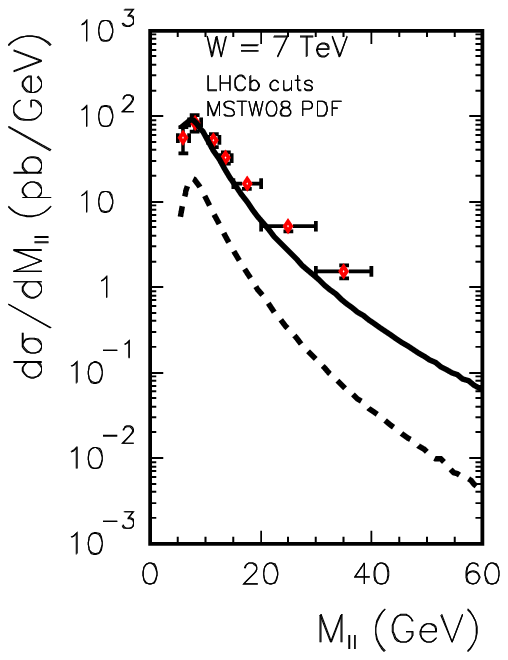}
\includegraphics[width=8cm]{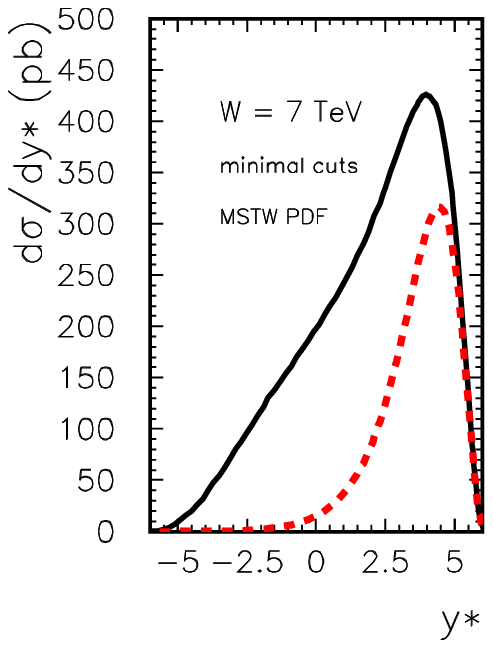}
  \caption{\label{fig:lhcb_dsig_dMll_second}
Contributions of the second-side component for the LHCb kinematics: 
2 $< y_{+},y_{-} <$ 4.5, $k_{T+},k_{T-} >$ 3 GeV. KMR UGDF was used here.
Right panel: Distribution in rapidity of the dileptons
for $\sqrt{s}$ = 7 TeV and $k_{T+},k_{T-} >$ 3 GeV
for MSTW08 valence quark distributions and KMR UGDFs.
The dashed line is the contribution from valence quarks only.
}
\end{figure}
%------------------------------------------------------------------------

In Fig. \ref{fig:lhcb_dsig_dMll_second} we return to the LHCb kinematics. 
In the left panel we show again the distribution in dilepton invariant mass. Here
by the dashed line we show the contribution from the ``other side'' proton. We see
that such a spillover of dileptons emitted into the forward region of ``the other''
proton is not negligible. It seems to be generally neglected in dipole model 
calculations. 
In the right panel we show the rapidity distribution of the virtual photon. 
By the red dashed line we show the contribution from valence quarks of the 
``forward'' proton only. We see that within the rapidity coverage of LHCb
sea quarks are important.

\section{Conclusions}
In our recent paper \cite{Schafer:2016qmk} we have considered Drell-Yan production of
dileptons in the forward rapidity region in a hybrid high-energy
approach. Corresponding formula for matrix element in the high-energy
approximation has been derived and presented in our recent paper.

Here we have shown some examples of differential cross
sections corresponding to recent experimental data for
low-mass dilepton production relevant for the LHCb and ATLAS
experiments.
Different UGDFs have been used in our calculations.

In contrast what was done in the literature, we have found
that both side contributions have to be included even for
the LHCb configuration.

We do not see clear hints of saturation at small $M_{ll}$.

\end{document}